\documentclass{article}
\usepackage{graphicx}
\usepackage{authblk}
\usepackage{verbatim}
\usepackage{array, tabularx}
\usepackage{subfigure}

\usepackage[a4paper, total={6.2in, 8in}]{geometry}
\usepackage{xr}
\usepackage[utf8]{inputenc}
\usepackage[T1]{fontenc}
\usepackage{amsthm}
\usepackage{amsmath}
\usepackage{amssymb}
\usepackage{siunitx}

\usepackage{url}
\usepackage{booktabs} 
\usepackage{xspace}
\usepackage{float}
\usepackage{multirow}
\usepackage{alltt}
\usepackage{color}
\usepackage{pifont}
\usepackage{makecell}
\usepackage{multicol}
\usepackage[super]{nth}

\newcommand\cparagraph[1]{\vspace{0.6mm}\noindent\textbf{#1}}

\let\origref\ref
\def\ref#1{\textnormal{\origref{#1}}}

\title{Fair Voting Methods as a Catalyst for Democratic Resilience: \\A Trilogy on Legitimacy, Impact and AI Safeguarding}

\author[1]{Evangelos Pournaras}

\affil[1]{School of Computer Science, University of Leeds, Leeds, UK. Email: e.pournaras@leeds.ac.uk}

\date{}
\begin{document}
	\maketitle
	
\begin{abstract}
This article shows how fair voting methods can be a catalyst for change in the way we make collective decisions, and how such change can promote long-awaited upgrades of democracy. Based on real-world evidence from democratic innovations in participatory budgeting, in Switzerland and beyond, I highlight a trilogy of key research results: Fair voting methods achieve to be (i) legitimacy incubator, (ii) novel impact accelerator and (iii) safeguard for risks of artificial intelligence (AI). Compared to majoritarian voting methods, combining expressive ballot formats (e.g. cumulative voting) with ballot aggregation methods that promote proportional representation (e.g. equal shares) results in more winners and higher (geographical) representation of citizens. Such fair voting methods are preferred and found fairer even by voters who do not win, while promoting stronger democratic values for citizens such as altruism and compromise. They also result in new resourceful ideas to put for voting, which are cost-effective and win, especially in areas of welfare, education and culture. Strikingly, fair voting methods are also more resilient to biases and inconsistencies of generative AI in emerging scenarios of AI voting assistance or AI representation of voters who would be likely to abstain. I also review the relevance of such upgrades for democracies in crisis, such as the one of Greece featured in the recent study of `Unmute Democracy'. Greek democracy can build stronger resilience via higher representation of citizens in democratic processes as well as democratic innovations in participation. Fair voting methods can be a catalyst for both endeavors.
\end{abstract}

	\section{Introduction}\label{sec:introduction}
	
	\cparagraph{"\emph{Government of the people, by the people, for the people,}} \emph{shall not perish from the Earth}" according to Abraham Lincoln in his iconic Gettysburg Address~\cite{Lincoln1980}. Yet, the unprecedented challenges that democracies face nowadays across the globe suggest the contrary. These are some longstanding challenges but also some highly timely and inevitably forthcoming. Scaling up participation in deliberative and voting processes remains an unsolved problem, often a result of failing to fairly represent citizens in collective decisions~\cite{Dryzek2019,Helbing2023,Pournaras2020}. The spread of misinformation, fake news and manipulative political advertising in social media influence elections results~\cite{Bovet2019,Flamino2023}. The ongoing penetration of generative artificial intelligence (AI) in democracy via chatbots~\cite{Vargiu2025,Kozlovai,Lin2025,Majumdar2024,Vassos2024} is expected to have colossal influence on voters and public opinion formation. How to reinvent (digital) democracy~\cite{Helbing2015} to address a fast-growing deficit in democratic resilience by such challenges? What upgrades are required to build up this democratic resilience, that is the preservation of participation, representation and integrity in collective decision-making processes? 
	
	\cparagraph{What this article is all about.} To answer these questions, I introduce another perspective for the foundations of our democracy, backed up by research and real-world evidence: \emph{collective decision making with fairer voting methods}. I argue how outdated majoritarian voting methods have left democracies highly vulnerable at different levels and unprepared to respond to new challenges such as those of representing diverse pluralistic societies or risks of using AI in democracy. Drawing on recent democratic innovations of applying fair voting methods in real world, in Switzerland and elsewhere, I highlight evidence from key research~\cite{Hausladen2024,Pournaras2025,Maharjan2024,Majumdar2024}, in particular a trilogy of democratic upgrades that involved at their core fair voting methods as (i) \emph{legitimacy incubator}, (ii) \emph{novel impact accelerator} and (iii) \emph{AI safeguard}. I show the striking evidence about the distinguishing qualities of fair voting methods in practice and I discuss the ongoing momentum building around those across the globe. Finally, I discuss the relevance of these insights for democracies in deep crisis such as the one of Greece. Through the prism of the recent study `\emph{Unmute Democracy}\footnotemark[18]' revealing in numbers the very poor experience of citizens with the functioning of democracy, I review the potential of democratic upgrades based on fair voting methods. 
	
	\cparagraph{Why we vote.} Voting is a cornerstone of democracy to reach a collective choice. Voting is simply a way for citizens to express their preferences independently, i.e. with an autonomy, without having to adhere or align to the opinion of someone else. It is also a way for citizens to reach a collective choice efficiently, while avoiding the struggle and the cost of going through the process of reaching a consensus with each other, which may turn into an impossibility task for large populations. Therefore, \emph{adhering to the rules of voting} is what makes collective decision making easier to operationalize in practice and at scale. 
	
	\cparagraph{Voting as a power delegation to the voting methods} When we vote, we accept: 
	\begin{itemize}
		\item \emph{The voting alternatives} put for voting from which choices are made;
		\item \emph{The voting rules}; these are: 
		\begin{enumerate}
			\item \emph{The ballot format} with which the preferences are expressed; 
			\item \emph{The ballot aggregation method} with which the preferences are brought together to calculate the final collective choice, i.e. the winners.
		\end{enumerate}
		\end{itemize}
	
	\noindent This means that voting is eventually a paramount (i) \emph{delegation of power} to the voting methods and (ii) \emph{compromise} we have to make for the freedom we preserve to express our preferences independently rather than jointly and in synergy with others, which apparently is a complex endeavor to purse at a large scale. This reflection is key and fundamental for the role of voting in democracy.
	
	 \cparagraph{Voting and majority: two sides of the same coin?} However, most people do not realize the importance and implications of (over-)relying on (existing) voting methods, in the sense of questioning whether they are appropriate for the priorities of a society. One reasons for this is the deep-rooted prevalence of the single-choice majoritarian voting. This is the method that dominates how we run our society and, paradoxically, is often associated with what democracy is all about. It turns out though that (i) majoritarian voting methods persistently fail to support resilient democratic societies~\cite{Emerson2020,Contucci2016,Pournaras2025} and (ii) alternative voting methods that are preferential and promote a more proportional representation of citizens' preferences are highly neglected~\cite{Emerson2020,Maharjan2024}. 
	 
	  \cparagraph{How majoritarian voting fails.} Single-choice majoritarian voting is the selection of one option among two or more alternatives. For instance, in national elections citizens usually choose a single party to vote among several ones, while in referenda citizens usually choose among two alternatives. The votes are summed up to choose one (single-winner voting) or more winners (multi-winner voting). These winners represent the most popular options (among the ones put for voting). Majoritarian voting is excessively prominent because of its \emph{simplicity} and because \emph{it can always produce an absolute majority winner} between two alternatives. If (i) this should be the ultimate aim and if (ii) two alternatives represent well the spectrum of choice,  it is something to (re)consider. Majoritarian voting is also often well-appreciated because it yields clear decision making (the one of the majority) and sovereign governance to implement the (majoritarian) will of the people\footnote{However, this is neither a necessary condition nor what practice always confirms~\cite{Emerson2020}. For instance, majority bonus systems may result in disproportional concentration of power, making it harder to keep representatives accountable.} On the other hand, majoritarian voting methods more and more fail to represent diverse societies accurately. They often threaten the cohesion of the society resulting in the so called `\emph{tyrannies of majorities}', where under-represented or minority groups are suppressed and left without any voice or power~\cite{Nyirkos2018}. It is harder to promote compromise (towards consensus) in majoritarian decision making, which is a necessary condition for a well-functioning democracy. Instead, priority is to elect a winning majority. 
	  
	  \cparagraph{Implication of over-relying on majoritarian voting.} Real-world practice shows that collective choice under majoritarian voting often has the following effects: opinions polarize, extremism develops and majority voting turns out to become a catalyst of populism~\cite{Emerson2020}. The rise of parties in parliaments with extreme ideological agenda~\cite{Georgiadou2018} or referenda that turn out to produce controversial outcomes and a divided society are often a result of relying on majoritarian voting. For example, the Brexit referendum in the UK in 2016 introduced a (binary) dilemma in the society that, to a large extent, was not accurately representing what people really aspired if the UK remained or left the European Union. Similarly, the Greek bailout referendum in 2015 still generates central political controversy in the Greek public sphere. Was it about the exit from Eurozone (and the European Union), a negotiation mandate for the Greek bailout, and if this was the case, was it actually honored or not? Were the political decisions made on the basis of the `No' result justified and legitimate? `Squeezing' complex decisions into a binary dilemma with the aim to satisfy a majority fundamentally undermines the capacity of the society to compromise, converge and discover alternative solutions that benefit more people. Consequently, low election turnouts and an overall abstaining of citizens from decision making and public matters is often attributed to implications behind majoritarian voting~\cite{Emerson2020,Ladner1999,Smith2018}. Although majoritarian voting is prominent for national elections and referenda, with often an embedding in constitutions~\cite{Mcginnis2001}, this critique has broader relevance and stronger applicability to other of facets democratic decision-making processes. Election of partisan leaders and parliamentary candidates~\cite{Mitchell2000}, prioritization of constitutional changes, legislation amendments and online petitioning~\cite{Navarrete2024}, decision making in citizens' assemblies and mini-publics~\cite{Beauvais2019,Setala2017}, electing representative committee members~\cite{Aziz2017} as well as participatory budgeting~\cite{Faliszewski2023,Maharjan2024,Yang2024} are some examples where alternatives to majority voting can have an impact. But are there alternative voting methods that are fairer and can improve democracy?

	\cparagraph{Fair voting methods as powerful alternative to majoritarian voting.} These methods rely on ballot formats that are: 
	\begin{enumerate}
		\item  \emph{Multi-choice and preferential ballot formats};
		\item \emph{Ballot aggregation methods to elect winners such that voters' preferences are proportionally represented}. 
		\end{enumerate} 
	
	\cparagraph{Fairer ballot formats.} Multi-choice voting allows a voter to support more than one option, leaving space to make compromises and reach consensus. A voter is more likely to be satisfied by a voting outcome if the voting method encourages such flexibility. In its simplest form, multi-choice voting is made with several approvals of the options in the ballot but preferential voting rules provide further flexibility for more expressive preferences. Preferential voting rules can represent more accurately the priorities of voters. For instance, a voter can \emph{rank} choices from highest to lowest preference\footnote{Ranked voting assumes the preferences of voters over the selected options are linear, i.e. the voter cannot directly express how much more (e.g. two, three or more times) an option is preferred over another one.}. Alternatively, each selected choice can be \emph{scored} (e.g. from 1 to 10). Moreover, there are several point-based systems that provide even more expressiveness to voters. For instance, with the \emph{modified Borda method}~\cite{Emerson2020} voters select a number of options to rank. The top-ranked option receives as many points as the number of options ranked. The second-ranked option receives one less, and so on. The more options ranked by the voter (i.e. the more flexible the voter is), the more points the top-ranked options receive, incentivizing in this way compromise. There are also voting methods that let voters \emph{distribute a number of points}. As an example, \emph{cumulative}~\cite{Cooper2007,Skowron2025} or \emph{quadratic}~\cite{Lalley2018,Wellings2023} voting let voters distribute a number of points. For instance, 5 points can be distributed to a maximum of 5 choices (one point each choice) and a minimum of 1 choice (all five points to one choice). A voter, who belongs to a minority group, may choose to raise a stronger voice for an option that is important for the group by assigning most of the points to this option. These methods, starting from simple multi-choice approval ballot formats to more expressive distributional systems provide a continuum of alternatives to majoritarian voting methods. Although it is well-known that these methods are vulnerable to manipulation as well and there is no perfect voting system that can satisfy a complete set of key fairness conditions (the so called Arrow's impossibility)~\cite{Arrow2010}, it important to distinguish that fair voting methods are a better fit for diverse and inclusive societies and, in practice, more useful to promote a consensus-driven collective decision making based on compromises, which is requirement for any well-functioning democracy. 
	
	\cparagraph{Fairer ballot aggregation.} Fairer ballot formats are not enough to reach more inclusive voting outcomes that represent broadly citizens' preferences. How preferences are aggregated is of utmost importance. Where the `\emph{standard}' majoritarian aggregation method (utilitarian greedy) prioritizes the most popular choices as winners\footnote{As long as constraints are satisfied, such as the number of available seats in the parliament for national elections or the available budget in participatory budgeting elections.}, alternative fair methods such as \emph{equal shares}~\cite{Peters2021} or the \emph{Phragmén's}~\cite{Brill2024} voting rule prioritize winners that achieve a more proportional representation of citizens' preferences\footnote{A difference of the two methods is the prioritization of the Phragmén’s method to balance representation between groups. On the contrary, the method of equal shares balances representation within groups, ensuring the representation of at least one voter from each group~\cite{Peters2020}.}. The method of equal shares is particularly applicable to \emph{participatory budgeting}. This is a bottom-up way to distribute the public budget of a city council by letting voters propose their own project ideas, for which they vote and often implement themselves. As project ideas come with different cost, popular ideas, which are usually expensive ones, are likely to overuse the budget with the standard majoritarian method, leaving other cost-effective alternatives under-represented~\cite{Maharjan2024}. In contrast, equal shares is a fairer approach by assigning to each voter the same decision power (i.e. equal share of the budget) to shape the voting outcome~\cite{Peters2021}. In practice and intuitively, this means that a popular and expensive project that would win under the standard majoritarian method, would loose under equal shares to elect instead several lower-cost projects, which, in combination, satisfy more voters. This striking effect provides a completely new and radical approach to what voting means and how it can yield more favorable outcomes for more people. Similarly to participatory budgeting where voters choose projects, these methods are also applicable for (costed) agendas outlining priority areas and positions (e.g. the agendas of candidates and parties). 
	
	\cparagraph{Breaking the vicious cycle of over-relying on majoritarian voting.} But if we have so many promising alternatives, why majoritarian voting methods keep prevailing in real-world practice? A long-standing vicious cycle of political stagnation prevents change: 
	
	\begin{enumerate}
		\item Until recently, evidence of how fair voting methods perform was mainly theoretical and academic-driven. No empirical insights, grounded to real-world evidence, was available. No blueprint of change existed. 
		\item  In the light of this deficit of evidence, fair voting methods have been often considered too complex to apply to real world, mainly a low-cost exercise of academic interest rather than a paradigm shift for democracy.
		\item This developed perception further reinforced the use of majoritarian voting methods, which politicians and the political system have by now mastered to manipulate to promote their own interests. Any alternative in this reinforced majoritarian reality is a threat to a well-established status quo and a `risk' that is hard to take in such a sensitive context, i.e. the narrative of "\emph{we do not play with democracy}'.
		\item This status quo further preserves the lack of evidence and the monopoly of majoritarian reality in the public sphere.
	\end{enumerate}
	
	\noindent But this vicious cycle (i) \textbf{can break} as I show based on recent real-world evidence and (ii) \textbf{should break} if we want our democracies to survive ongoing and forthcoming challenges, in particular, the new inevitable risks of using AI in democracy.

	\section{A Trilogy of Democratic Upgrades using Fair Voting Methods}\label{sec:trilogy}
	
	\cparagraph{Fair voting methods for legitimacy, impact and AI safeguarding.} Figure~\ref{fig:representation} illustrates a trilogy of democratic upgrades as a result of applying fair voting methods in real-world, in particular for participatory budgeting, but also beyond this. These upgrades demonstrates that fair voting methods serve as:
	
		\begin{figure}[!htb]
		\centering
		\includegraphics[width=1.0\textwidth]{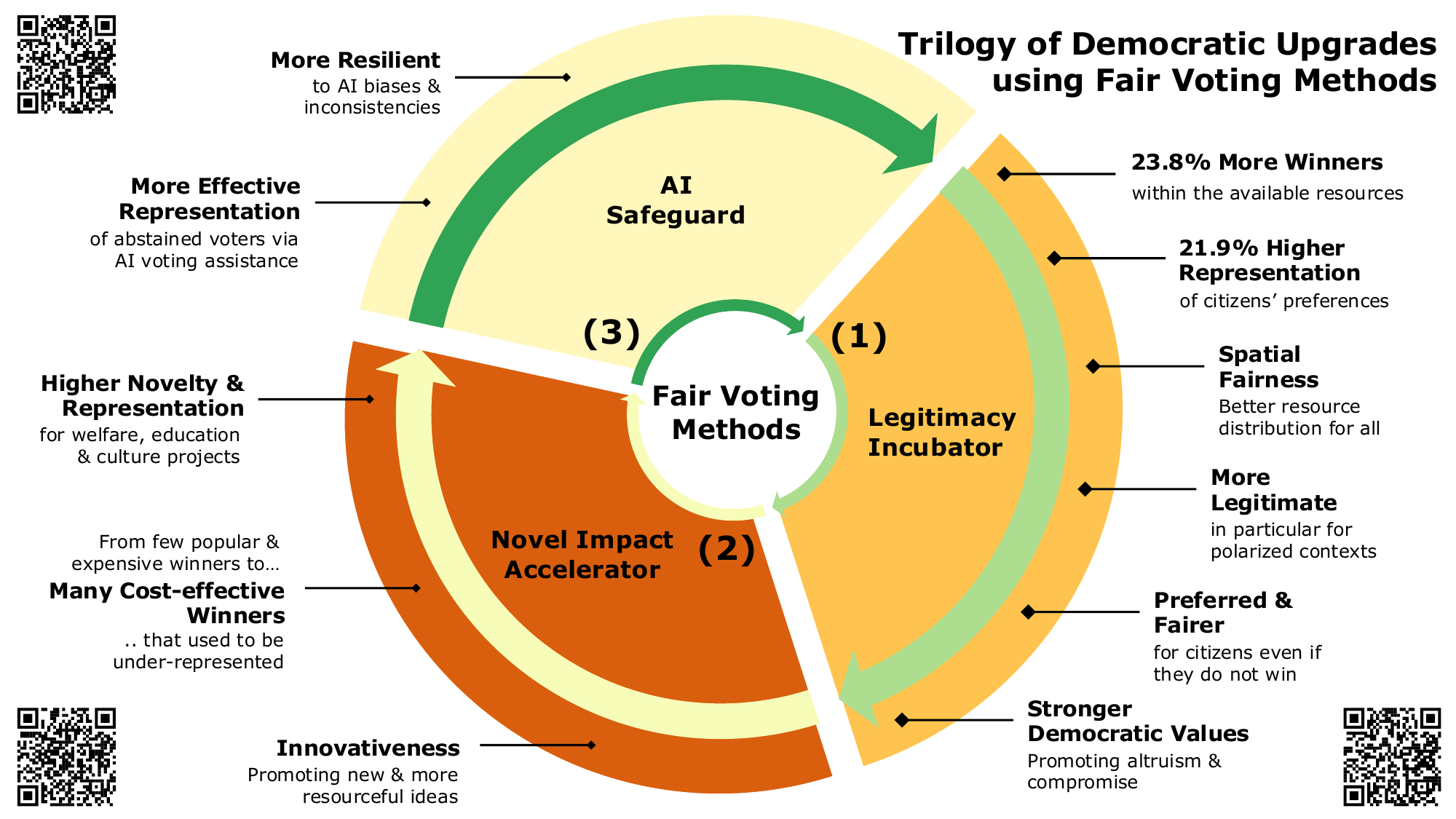}
		\caption{\textbf{A trilogy of democratic upgrades using fair voting methods}. Evidence from real-world democratic innovations demonstrate fair voting methods as a (i) legitimacy incubator~\cite{Pournaras2025}, (ii) novel impact accelerator~\cite{Maharjan2024} and (iii) AI safeguard~\cite{Majumdar2024}.  Fair voting methods outperform majoritarian ones by providing higher representation levels of citizens' preferences with more winners, higher spatial fairness, while citizens prefer and find them fairer, exhibiting stronger democratic values. Fair voting methods also involve more innovative and resourceful ideas put for voting, resulting in novel  and cost-effective winning projects that better represent welfare, education and culture that have been so far under-represented. Fair voting methods are more resilient to AI biases and inconsistencies, providing a more effective representation of abstained voters who receive AI voting assistance to participate.}
		\label{fig:representation}
	\end{figure}
	
	\begin{enumerate}
	\item \textbf{Legitimacy incubator}: Adopting fair voting methods yields voting outcomes that are more legitimate, in particular for polarized contexts~\cite{Hausladen2024}; these are outcomes with more winners and better (geographical) representation of citizens; they align with the preferences of more people even if those are losers as a result of building up stronger democratic values~\cite{Pournaras2025,Hanggli2024}. 
	
	\item \textbf{Novel impact accelerator}: Adopting fair voting methods yields voting outcomes with novel and resourceful winners that used to under-represented and make a most cost-effective use of resources~\cite{Maharjan2024}. 
	
	\item \textbf{AI safeguard}: Adopting fair voting methods in scenarios of AI voting assistance or representation of abstained voters yields voting outcomes that have less biases and inconsistencies. Fair voting methods create democratic resilience against low turnouts and risks by (inevitable) involvement of AI in voting processes~\cite{Majumdar2024}. 
	 
	\end{enumerate}
	
	The evidence of these striking democratic innovations is based on the milestone of Aarau in Switzerland, where fair voting methods are thoroughly evaluated in a real-world setting~\cite{Pournaras2025}.

	\section{A Blueprint of Upgrading Democracies with Fairer Voting Methods}\label{sec:Aarau}
	
	\cparagraph{The missing real-world evidence on fair voting methods.} Two fair voting methods were rigorously tested\footnote{The city of Wieliczka in Poland also ran in 2023 the participatory budgeting process of `\emph{Green Million}' using the method of equal shares.} for the first time in the city of Aarau in Switzerland in 2023 in the context of a participatory budgeting process, the \emph{City Idea}\footnote{Available at \url{https://www.stadtidee.aarau.ch} (last accessed: December 2025).}  (`\emph{Stadtidee}'). These methods are the (i) \emph{cumulative voting}, with the distribution of 10 points to at least three projects\footnote{Requiring at least three projects to select is introduced to eliminate severe strategic voting, in which a voters would assign all points to a single project. Apparently, among 33 project alternatives and with enough funding to implement several ones, it would be unfair to monopolize the voting power of voters to single projects, which in turn would emulate single-choice majoritarian voting. Nevertheless, voters can still distribute a large number of points to their most preferred project, i.e. 8, 1, 1. We observe that in practice this does not happen often and voters preserve a plurality of preferences.} and the (ii)  \emph{equal shares}\footnote{Equal shares is used with the \emph{Add1U} completion method that makes sure most of the budget is allocated after prioritizing a proportional representation of citizen preferences~\cite{Papasotiropoulos2025,Faliszewski2023}.}. Therefore, a fairer approach is introduced for \emph{both the ballot format and the ballot aggregation}. The participatory budgeting process of Aarau is the first of its kind, not only for testing fair voting methods in real-world, but also because of its robust evidence based on a rigorous experimental design and several innovations in citizens' participation~\cite{Pournaras2025}. 
	
	\cparagraph{A rigorous systematic approach to prove how fair voting methods thrive in real world.} Two survey studies were conducted right before and after voting, linking participants at each stage of participation. With this approach, different voting outcomes reached by different voting methods are compared to identify how well voters are represented at each case~\cite{Pournaras2020}. Ultimately, we were able to learn which voting methods citizens prefer and find fairer to use. We were also able to determine which human factors causally explain the choice of the voting method as the preferred and fair one. This systematic approach, (grounded on causal inference), is distinguished by other existing approaches that are more qualitative, descriptive and often anecdotal, with the risk of generating spurious correlations and contentious evidence. 
	
	\cparagraph{Fair voting methods can attract significant and diverse participation.} The participatory budgeting process in Aarau engaged 1,703 voters, including immigrants and children over 15 years old to distribute a public budget of 50,000 CHF. Although the turnout was not at the level of national elections or referenda, this is still a significant number for such processes and a small city of 14,310 eligible voters compared to other participatory budgeting processes\footnote{For instance, Zurich engaged a similar number of voters in 2021 although the city itself and the available budget to distribute are much larger~\cite{ZurichPBReport}.}. Moreover, 3,593 voters participated in the survey before voting and 808 after voting. Citizens proposed more than 161 project ideas out of which 33 were put for voting. Citizens were aware about the voting methods that would be used when they proposed their project ideas. Citizens voted using the open-source platform of Stanford Participatory Budgeting~\cite{Wellings2023}. They proved their eligibility to vote in a privacy-preserving way with (a hash of) their social security number. 
	
	\cparagraph{Beyond Aarau: generalizing real-world evidence of fair voting methods.} Applying fair voting methods in real world opens up opportunities for several novel spatio-temporal comparisons of voting outcomes that were not so far possible. These are about the outcomes of: 
	
	\begin{itemize}
		\item \textbf{Comparing local}: The actual use of the fair voting method vs. the hypothetical use of the standard majoritarian method in Aarau. 
		\item \textbf{Comparing global}: The actual  use of the fair voting method in Aarau vs. the actual use of the standard majoritarian method in past elections. 
		\item \textbf{Confirm on past evidence}: The hypothetical use of the fair voting method vs. the actual use of the standard majoritarian method in past elections. 
		\item \textbf{Confirm on future evidence}: The actual use of the fair voting methods in Aarau vs. the actual use of the fair voting method in follow up elections.
	\end{itemize}
	
	\noindent Detailed data of past participatory budgeting elections are available in the open Participatory Budgeting Library\footnote{Available at \url{https://pabulib.org/} (last accessed: December 2025)} (Pabulib). Follow up evidence of applying equal shares in real world includes the cities\footnote{Available at \url{https://equalshares.net} (last accessed: December 2025).} of Świecie in Poland, Assen in the Netherlands and Winterhur in Switzerland. The comparisons based on these real-world data constitute a solid basis for a robust and complete evidence of the transforming character of fair voting methods. 

	The trilogy of democratic upgrades based on fair voting methods is illustrated in the next three sections. It unfolds based on three extensive studies~\cite{Pournaras2025,Maharjan2024,Majumdar2024} with a detailed methodology that is not within the scope of this article. As such, I focus here on linking these studies as a continuum of democratic upgrades, highlighting the key findings and take-away messages.

	\section{1. Fair Voting Methods as Legitimacy Incubator}\label{sec:legitimacy} 
	
	This study~\cite{Pournaras2025} shows that, strikingly, 23.8\% more winning projects are achieved using fair voting methods in participatory budgeting as compared with the standard majoritarian method, \emph{using the same budget}. In the case of Aarau, 17/33 projects were funded instead of only 7/33 that would be funded with the standard majoritarian method. 
	
	How this is possible, one may wonder. Fair compromises are a built-in design element of fair voting methods. Without explicitly instructed to do so, they substituted in the winning set the third and sixth most popular projects that were expensive ones with 12 other more cost-effective projects such that they satisfy the preferences of more citizens: 75\% vs. 50\%.
	
	This innovative effect of fair voting methods has some remarkable implications for citizens: the representation of their preferences increases by 21.9\% on average; and spatial fairness increases as well by improving the geographical distribution of the implemented projects over the different districts of the city. For instance, 5 districts of Aarau end having more winning projects with fair voting methods compared to the results of the standard majoritarian method. 
	
	These profound qualities build up a strong legitimacy as the ultimate endeavor of democratic innovations. If citizens, who may also be losers (their preferred projects are not selected) can shift in accepting such voting outcomes as the ones they prefer and the ones they find fairer over the outcomes of the standard majoritarian voting method, this provides strong evidence that fair voting methods are more legitimate\footnote{This is also confirmed in a separate study, in particular for decisions made in polarized contexts~\cite{Hausladen2024}.}~\cite{Weatherford1992,Hanggli2024}. This hypothesis is rigorously tested for the first time in Aarau by tracing citizens' choice over the fair vs. majoritarian voting method before and after voting. Results show that 26\% and 44\% more citizens prefer and find fairer the new tested voting methods after voting.
	
	What explains this citizens' shift to fair voting methods? Are there any personal human traits that causally explain why citizens prefer and find fairer these methods? Strikingly, democratic values such as \emph{altruism} and \emph{compromise} explain the shift of citizens to fair voting methods. This suggests that adopting these methods, which exhibit strong legitimacy, comes along with building stronger democratic values in society.

	\section{2. Fair Voting Methods as Novel Impact Accelerator}\label{sec:impact}
	
	Fair voting methods in participatory budgeting result in different (i) \emph{proposed} and (ii) \emph{winning} projects. What is the anticipated impact by implementing such different projects? Which impact areas\footnote{The projects of 345 participatory budgeting elections~\cite{Maharjan2024} in Pabulib~\cite{Faliszewski2023} come with the following labels that characterize impact areas that these projects cover: education, public transit, health, welfare, public space, urban greenery, culture, sport and environmental protection.} of society are expected to benefit the most? So far, the participatory budgeting practice shows an over-representation for infrastructural and sustainable development projects that are usually expensive and over-represented~\cite{Maharjan2024}. Does this change when using fair voting methods? 
		
	As the citizens who participate in the project ideation phase know that the higher the cost of projects they propose for voting, the more voting support they would need to guarantee, they also come with more efficient and cost-effective project ideas, avoiding projects that are costly. This can discourage corruption and over-reliance on financial interests. The adoption of equal shares in Aarau confirms that a proposed project in Aarau is on average 48.4\% lower in cost\footnote{Measured by the mean cost share of the total budget.} than a typical project proposed under the standard majoritarian method. This also suggests a shift to novel proposed projects of different nature that are also likely to win as fair voting methods result in more winners (see previous section). 
	
	Fair voting methods result in novel winning projects with higher diversity on welfare, education and culture that used to be under-represented with the standard majoritarian voting method. In contrast, infrastructural and sustainable development projects loose representation in the winners due to their inherently higher cost. This can be mitigated by breaking down such large projects into smaller ones that can be funded over multiple participatory budgeting rounds, keeping in such a way large and risky projects more accountable to citizens. For instance, developing a network of bike lanes can funded at multiple stages, incrementally expanding the network in a grassroots fashion. This approach could reduce corruption and preserve citizens in the loop when implementing projects with long-term impact for their lives. 
	
	\section{3. Fair Voting Methods as Safeguard against AI Biases and Inconsistencies}\label{sec:safeguard}
	
	Despite the outstanding effectiveness of fair voting methods in participatory budgeting processes, scaling up participation in such processes remains a long-standing challenge. Turnouts in such processes are usually low~\cite{Sintomer2008,Bartocci2023}. Although a significant number of 1,703 citizens voted in Aarau out of 14,310 eligible voters, participation is lower than other collective decision-making processes such as referenda or national elections. However, even for national and supranational elections, the participation\footnote{International IDEA Voter turnout Database \url{https://www.idea.int} (last accessed: December 2025).} level is on average 61\% across the globe in 2024. 
	
	Abstaining voters may often have legitimate reasons to abstain and this stand often gives a powerful message for democratic change. However, abstaining because of low engagement of citizens with democratic processes~\cite{Hylton2023,Lane2017}, low digital literacy~\cite{Aichholzer2020,Vassil2011} or low trust to democratic processes~\cite{Wang2016,Belanger2017} may result in inaccurate voting outcomes~\cite{Li2025}, i.e. abstaining can change the voting result that would be reached if the abstained voters participated. 
	
	Digital participation and AI voting assistance emerge as an alternative to scale up participation and recover from such inaccurate election results. For instance, in Estonia 99\% of citizens have a digital identity and can electronically vote since 2005~\cite{Kattel2019}. As AI becomes more and more pervasive and ubiquitous, interfacing technology for digital participation with AI representatives\footnote{An AI representative is in practice an intelligent software agent (e.g. chatbot in a web or smart phone app) that can provide voting advice to a voter~\cite{Vassos2024} (e.g. summarizing the political agendas of parties)or even make (approved) choices on voter's behalf based on preferences and past choices. The motivation for the latter scenario is scaling up participation into a large number of decisions that is now exercised by elected representatives in the parliament, for instance.} is a plausible scenario, which may turn out to be a technological inevitable in the future, rather than what we have broadly accepted to pursue~\cite{Vargiu2025,Kozlovai,Lin2025}. The question that arises is how the integrity of elections is influenced by AI representatives of abstained voters in such a scenario. Can AI representatives actually recover the lost integrity of elections that experience low turnouts?  Or do they instead undermine the integrity of elections further because of biases and inconsistencies that AI algorithms possess as well as their manipulative capabilities? And what role fair voting methods can play to mitigate such influences? 
	
	To address these questions, more than 50,000 AI voting personas were constructed that estimate voting choices of abstained voters based on real-world profiles of voters in 363 scenarios of participatory budgeting and national elections. These AI representatives are based on several large language models\footnote{Gemini 1.5 Flash, Deepseek R1, Llama, GPT 3.0, 3.5, and 4o mini} and predictive machine learning. The consistency of the voting outcomes under the fair and the standard majoritarian voting method is assessed under (i) varied levels of abstained voters and (ii) varied levels of AI representation of these abstained voters. 
	
	The results of this study~\cite{Majumdar2024} show that fair voting methods are more resilient of AI biases and inconsistencies. The consistency of the voting outcomes reached using fair voting methods is not significantly influenced by inaccurate voting estimates. In contrast, the consistency of the voting outcomes reached with the standard majoritarian method can be much more vulnerable to inaccurate estimates, with several cognitive biases explaining such inconsistencies between the choice of human voters and their AI representatives. Moreover, AI representation is more effective for abstained voters rather than any arbitrary voter. This suggests that participation is irreplaceable and it remains the ultimate safeguard for the quality of democracy in the era of AI. Strikingly, fair voting methods are more effective than the standard majoritarian method to preserve winning projects in the voting outcomes that would not be there without the AI representation of abstained voters. 
	
	\section{The Case of Greece: Fair Voting Methods to Unmute Democracy}\label{sec:unmute-democracy}
	
	\cparagraph{Why Greece.} The capacity of fair voting methods to empower democratic upgrades is reviewed here in the context of Greece and the recent study of `\emph{Unmute Democroracy}'\footnote{Available at \url{https://eteron.org/research/unmute-democracy-research/}. The survey study is conducted by Aboutpeople (\url{https://aboutpeople.global}) on behalf of Eteron (last accessed: December 2025).} of Eteron\footnote{Available at \url{https://eteron.org} (last accessed: December 2025)} and Vouliwatch\footnote{Available at \url{https://vouliwatch.gr} (last accessed: December 2025)}. The study is based on a survey conducted in September 2025 using a cross-national sample of 1,876 citizens in Greece. It aims to assess citizens' perception about the functioning democracy in Greece nowadays. It captures threats for democracy, possible mitigation actions at an institutional level, types of political participation that can improve the quality of democracy and empower citizens' participation in public matters. Greece is chosen as a case study due to its strong historic association to democracy, which is though heavily challenged nowadays at different levels (as the study itself confirms). Therefore, this is a highly-relevant context to assess how democratic upgrades based on fair voting methods can rebuild a long-awaited democratic resilience to some long-standing and other emerging challenges for democracy in Greece and beyond.
	
	\cparagraph{The imperative of democratic upgrades.} The key take-away message from this study confirms the paramount democratic deficit experienced by Greek citizens. The functioning of democracy in Greece is assessed with 3.9/10 (3.2/10 for young people in the age range of 17-34). This suggests the urgent need for democratic upgrades that can restore trust of people to democracy. Citizens identify a significant gap (80.5\%) for politicians and parties in acquiring what citizens believe about important public matters. Thus, voting methods could play a more significant role so that, when used in any of these contexts, they represent citizens more fairly. Strikingly, the quality of democracy would improve by 28.4\% more if politicians and parties listened citizens’ opinion compared to the one of experts and technocrats (90.1\% vs. 61.7\%). A question that arises here is whether the means that governments have available to address such opinions are so effective for citizens as they are for experts and technocrats. The engagement of experts and technocrats is prominent but not always transparent. Often, it does not align with the citizens' expectations and interests for which decisions are made. On the contrary, citizens opinions are mainly expressed in national elections every four years using majoritarian voting methods and electoral systems with limited representation power. This dual misalignment undermines democracy. Instead, fair voting methods could be used to align the voice of experts and technocrats with an upgraded voice of citizens. For instance, the participatory budgeting blueprint of Aarau demonstrates a way to engage citizens in meeting net-zero targers set by policymakers by providing resourceful alternatives to citizens to choose from, in a way that promotes fair compromises for the benefit of more people~\cite{Pournaras2025,Maharjan2024}. 
	
 	\cparagraph{Opportunities and challenges for upgrading Greek democracy.} The democratic capital of citizens' participation and empowerment in collective decision making is strongly acknowledged along with the challenge to develop it. Participants of the study considered the option of having the right to propose to parliament the passage, repeal or amendment of a law by collecting 500,000 signatures. Although they find this participation as an improvement in democracy by 50.5\%, they also find it to a large extent inapplicable (24.4\%). They also find it likely to sign a petition by 79.4\% but very unlikely to initiate one (9.8\%). At a more local level, the participation in local citizens' assemblies is supported by 58.5\%, while still a large share of the population (38.8\%) does not find it likely to participate. A simple proportional representation system to better represent citizens' voices (50.9\% ) is preferred over a proportional representation systems that provide bonus of seats in the parliament for stronger governance (31.4\%). With the exception of sortition for electing members of the parliament (52.8\% against), citizens strongly support democratic innovations such as participatory budgeting (70.8\%), digital participation via platforms and apps (73.8\%), e-voting in elections (69.6\%) and even referenda (80.6\%) at both local/national level and also despite the significant controversy of the Greek bailout referendum in 2015. These results suggest that citizens preserve the will to participate in democracy, however, they may also find it futile if their voice has no impact, especially as moving from a local to a national level. As such, the participation blueprint of Aarau based on fair voting methods is particularly timely and relevant for Greek citizens as it meets and fits with their expectations of how to improve the quality of democracy. 
	
	\cparagraph{The paradox of democratic upgrades.} Despite the potential to unleash the democratic capital in the Greek society, the study suggests several factors that can hinder democratic innovations. First and foremost, low civic participation and engagement (19.1\%) is not found such key threat for the functioning of democracy as the corruption of the political system (59.4\%), lack of justice (57.6\%), political/economic influence on media and their quality (32.3\%) as well as economic inequality (24.4\%). Similarly, while means of direct democracy are strongly supported (33.5\%), citizens see some more important obstacles that hinder the quality of democracy such as (i) the influence of large economic interests (55.7\%) that dominate policy-making in Greece and (ii) ensuring the separation of powers (40.6\%). To what extend are those top-2 factors the necessary conditions to initiate democratic upgrades? Or could democratic upgrades become themselves the necessary conditions to transform democracy as a whole? This paradoxical relationship is hard to disentangle and also has other facets within the study. For instance, the non-representative electoral systems are not prominently identified as a threat for the functioning of democracy (5.3\%). Institutionalizing internal party elections for the nomination of parliamentary candidates is the second lowest chosen intervention (14.8\%). Citizens' participation via voting, deliberation and mobilizations is not so important as institutions and justice that keep governments accountable (23.7\% vs. 54.7\%). And strikingly, weak representation of citizens within political parties is not so major problem (24.5\%) as the lack of competent and honest members (40.2\%) or the dependence on financial interests (30.6\%). This prioritization suggests an expectation for a required top-down change to improve democracy. On the contrary, the blueprint of Aarau suggests that this `chicken-and-egg' problem can be easier to approach at a local level first, where the influence of large economic interests and the separation of powers may not prohibit democratic innovations. A more gradual scaling up and scaling wide of democratic upgrades is an alternative way forward for change. 
	
	\cparagraph{The role of AI.} Contemporary threats to democracy by the use of AI are not prominent among citizens. Misinformation and the spread of fake news are recognized as one of the top-3 threats by 16\% of the participants, while dangers by the use of AI is supported by only 2.5\%. Instead, the use of AI for detecting and reducing corruption is supported by 72.6\% of participants. In other words, citizens seem to identify much more opportunities than risks to AI. However, there are objective (and inevitable) risks in using AI for scaling up and establishing actions that citizens strongly support such as more direct democracy, participatory initiatives, referenda, e-voting and citizens' mobilization via social media. Fair voting methods are a safeguard that cannot be neglected to make sure that using AI meets these expectations of citizens as the evidence from recent research suggests~\cite{Majumdar2024}. 

	\cparagraph{Take-away messages.} The review of how  democratic upgrades based on fair voting methods can rebuild a long-awaited democratic resilience in Greece is summarized in the following four points:

\begin{enumerate}
	\item Where fair voting methods could promote an inclusive representation of Greek citizens to improve democracy, they fail to play a (significant) role in practice, which exacerbates the large democratic backsliding for the functioning of democracy on citizens' eyes. 
	
	\item Fair voting methods align and can promote citizens' priorities towards more proportional representation systems as well as strongly supported democratic innovations such as participatory budgeting, digital participation, e-voting and referenda. 
	
	\item Fair voting methods suggest a bottom-up way to improve democracy, which does not always align with citizens' priorities on top-down interventions for a fairer and more independent juridical system, lower corruption and more competent leaders with ethos. 
	
	\item Fair voting methods as a safeguard to AI risks is highly relevant for Greek citizens as they show low awareness of these risks, while they preserve high expectations from using AI to improve quality of democracy. 
\end{enumerate}

From Switzerland, to Greece, these observations, in the light of recent research evidence, can be used as an agenda for further discourse and knowledge exchange on the role of fair voting methods for the future of democracies across the globe. 

\section{Conclusion and Future Perspective}\label{sec:conclusion} 
	
	To conclude, the majoritarian `obsession' proves to be outdated and insufficient to respond to old and new challenges of democracy. Over-relying on majority voting for collective decisions can further exacerbate low participation, low voting turnouts, extremism, polarization and further divide of the society. Consequently, the advent of AI within a majoritarian democracy provides further opportunities to automate and scale up all these threats in unprecedented ways we have not witnessed so far. As a response and an alternative, I outline a trilogy of real-world democratic upgrades based on fair voting methods. These upgrades demonstrate fair voting methods as (i) \emph{legitmacy incubator}, (ii) \emph{novel impact accelerator} and (iii) \emph{AI safeguard}. Bringing these three facets of evidence together is a statement of how fair voting methods can break a longstanding vicious cycle of democratic stagnation. As a summary of this evidence, compared to the standard majoritarian rules, fair voting methods yield~\cite{Pournaras2025,Hausladen2024,Maharjan2024,Majumdar2024}: 
	
	\begin{enumerate}
		\item More winners within the available resources~\cite{Pournaras2025,Maharjan2024};
		\item Higher representation of citizens' preferences~\cite{Pournaras2025};
		\item Higher spatial fairness with better representation for the periphery of cities~\cite{Pournaras2025}; 
		\item Higher legitimacy, especially in polarized contexts. They are preferred and found fairer even for the voters who do not win~\cite{Pournaras2025,Hausladen2024};
		\item Stronger democratic values to promote for citizens~\cite{Pournaras2025};
		\item Innovativeness by promoting new and more resourceful ideas~\cite{Maharjan2024};
		\item Winners that are more cost-effective and well-representing~\cite{Pournaras2025,Maharjan2024};
		\item Novel and better represented project ideas for welfare, education and culture~\cite{Maharjan2024};
		\item More resilience against AI biases and inconsistencies~\cite{Majumdar2024};
		\item More effective representation of abstained voters via AI voting assistance~\cite{Majumdar2024}.
		\end{enumerate}
		
	Although this real-world evidence relies to a large extent on participatory budgeting experience that has a more local scope, a growing momentum shows that fair voting methods can scale up and wide. For instance, I identified several follow up participatory budgeting processes that further adopt the fair voting methods of Aarau. Equal shares is likely to be the next golden standard for aggregating ballots in participatory budgeting. In parallel, the paradigm of participatory budgeting based on fair voting methods is explored in other budgeting processes. For instance, the distribution of research funding\footnote{For instance the ongoing project: \emph{Embedding EDI in the distribution of research funding: an AI-assisted collective intelligence approach} funded by the AI for Collective Intelligence Hub: \url{https://ai4ci.ac.uk/funded-projects/} (last accessed: December 2025).}, crowdfunding, the distribution of funds for tackling climate change and meeting net-zero targets as well as the governance of blockchain software communities are further opportunities to explore the impact of fair voting methods. 
	
	The trilogy of democratic upgrades based on fair voting methods can be particularly valuable as a blueprint of rebuilding democratic resilience, particularly for democracies in crisis such as the one of Greece. Greek citizens appreciate approaches for direct democracy, including participatory budgeting, e-voting and referenda. However, the major challenges they face with democracy are systemic and related to corruption, justice, the economic and political power of media as well as the leadership of parties. The participatory budgeting approach of Aarau based on fair voting methods motivates an alternative governance framework, that can inform policy at all levels: \emph{More novel and inclusive decisions to make among more resourceful options, to achieve better (geographic) representation with more winners, while building up democratic resilience against corruption, economic monopolies and risks of AI.}

	\section*{Acknowledgments}
	
	Evangelos Pournaras is supported by a UKRI Future Leaders Fellowship (MR\-/W009560\-/1): `\emph{Digitally Assisted Collective Governance of Smart City Commons--ARTIO}'. The underlying research is done with support of several people including among others (in alphabetical order): Edith Elkind, Regula Hänggli, Dirk Helbing, Fatemeh B. Heravan, Sajan Maharjan, Srijoni Majumdar, Joshua C. Yang, Thomas Wellings. 
	
	\bibliography{bibliography}
	
	\bibliographystyle{naturemag}

\end{document}